\documentclass[aps,prl,twocolumn,showpacs,floatfix,superscriptaddress]{revtex4}
\usepackage{graphicx}
\usepackage{graphics}
\usepackage{amsmath}
\usepackage[usenames]{color}

\newcommand{\be}{\begin{equation}}
\newcommand{\ee}{\end{equation}}
\newcommand{\bea}{\begin{eqnarray}}
\newcommand{\eea}{\end{eqnarray}}

\begin{document}

\title{Energy decay in three-dimensional freely cooling granular gas}
\author{Sudhir N. Pathak}
\email{sudhirnp@imsc.res.in}
\affiliation{The Institute of Mathematical Sciences, CIT Campus, Taramani, 
Chennai-600013, India}
\author{Zahera Jabeen}
\email{zjabeen@umich.edu}
\affiliation{ Department of Physics, University of Michigan, Ann Arbor, 
MI 48109-1040, USA}
\author{Dibyendu Das}
\email{dibyendu@phy.iitb.ac.in}
\affiliation{Department of Physics, Indian Institute of Technology,
Bombay, Powai, Mumbai-400 076, India}
\author{R. Rajesh}
\email{rrajesh@imsc.res.in}
\affiliation{The Institute of Mathematical Sciences, CIT Campus, Taramani, 
Chennai-600013, India}
  
\date{\today}

\begin{abstract}
The kinetic energy of a freely cooling granular gas decreases as a
power law $t^{-\theta}$ at large times $t$. Two theoretical
conjectures exist for the exponent $\theta$. One based on ballistic
aggregation of compact spherical aggregates predicts $\theta=
2d/(d+2)$ in $d$ dimensions. The other based on Burgers equation describing anisotropic,
extended clusters predicts $\theta=d/2$  when $2\le d \le 4$. 
We do extensive simulations
in three dimensions to find that while $\theta$ is as predicted by
ballistic aggregation, the cluster statistics and velocity
distribution differ from it. Thus, the freely cooling
granular gas fits to neither  the
ballistic aggregation or a Burgers
equation description.
\end{abstract}  
\maketitle

The freely cooling granular gas, a collection of ballistically moving
inelastic particles with no external source of energy, has been used to 
describe dynamics of granular 
materials~\cite{aranson2006,granulargases,granulardynamics}, large scale 
structure formation in the universe~\cite{shandarin1989} and
geophysical flows~\cite{campbell}.
It is also of interest as a 
system far from equilibrium, limiting cases being amenable to exact
analysis~\cite{frachebourg,sabhapandit2009}, 
has close connection to the well studied Burgers
equation~\cite{frachebourg,frachebourg2000,olegburger,kida,supravat2011}, and
is an example of an ordering system showing non-trivial coarsening
behavior~\cite{puri,shinde2007,shinde2009,shinde2011}.
Of primary interest is clustering of particles due to inelastic collisions and 
the temporal evolution of the kinetic energy $E(t)$ at large times. 

At initial times, particles remain homogeneously distributed and 
kinetic theory predicts that 
$E(t)$ decreases as $(1+t/t_0)^{-2}$ (Haff's law) 
where the time scale $t_0 \propto (1-r^2)^{-1}$ for constant coefficient
of restitution $r$~\cite{haff}. At later times, this regime is
destabilized by  long wavelength fluctuations into an
inhomogeneous cooling regime dominated by clustering of
particles~\cite{goldhirsch1993,mcnamara_pre,efrati2005}. 
In this latter regime, $E(t)$ no 
longer obeys Haff's law but decreases as a power law 
$t^{-\theta}$, where $\theta$ depends only on  
dimension $d$~\cite{bennaim1999,nie2002}. Direct experiments on 
inelastic particles under levitation \cite{maa} or in microgravity
\cite{haffexperi1,Grasselli2009}  confirm Haff's law. However, being
limited by small number of particles and short times, 
they do
not probe the inhomogeneous regime giving no information about $\theta$.

Different theories predict  different values of $\theta$.
The extension of kinetic theory into the inhomogeneous cooling regime 
using mode coupling methods leads to 
$E(\tau)\sim \tau^{-d/2}$, where the relation between the average number of 
collisions per particle $\tau$ and time $t$ is unclear~\cite{britoernst}.  
This result agrees with simulations for near-elastic ($r\approx 1$)
gases, but fails for large times 
and strongly inelastic $(r \ll 1$) gases~\cite{britoernst}.
Any theory involving perturbing about the elastic limit $r=1$ is unlikely 
to succeed 
since extensive simulations in one~\cite{bennaim1999} and two~\cite{nie2002}
dimensions show that for any $r<1$, the system is akin to a {\it sticky gas}
($r\rightarrow 0$), such that colliding particles stick and form
aggregates.

If it is assumed that the aggregates are compact spherical objects,
then the sticky limit corresponds to the well studied ballistic
aggregation model (BA) (see Ref.~\cite{leyvraz2005} for a review). 
For BA in the dilute limit and the mean field 
assumption of uncorrelated aggregate
velocities, scaling arguments lead to $\theta_{BA}^{mf}= 2 d/(d+2)$ and the
presence of a growing length
scale ${\mathcal L}_t \sim t^{1/z_{BA}^{mf}}$ with 
$z_{BA}^{mf}= (d+2)/2$~\cite{carnevale}. 
In one dimension, BA is exactly solvable and
$\theta_{BA}=\theta_{BA}^{mf}$~\cite{frachebourg,frachebourg2000}. 
However, in two dimensions 
and for dilute systems, it has been shown that
$\theta_{BA}^{mf}$ is smaller than the numerically obtained 
$\theta_{BA}$ by $17$\% because of 
strong velocity correlations between colliding 
aggregates~\cite{trizac1,trizac4}. 

The sticky limit has also been conjectured~\cite{bennaim1999,nie2002} 
to be describable by Burgers-like equation (BE)~\cite{burgersbook}. This
mapping is exact in one dimension~\cite{kida} and heuristic in two and
higher dimensions~\cite{nie2002}, and leads to 
$\theta_{BE}=2/3$ in $d=1$,  $\theta_{BE}=d/2$ for $2\leq d\leq
4$, and $\theta_{BE}=2$ for
$d>4$~\cite{shandarin1989,esipov1993,esipov1994}.

The exponents $\theta_{BA}^{mf}$ and $\theta_{BE}$
coincide with each other in one and two dimensions and also with 
numerical estimates of $\theta$ for the freely cooling 
granular gas in these dimensions~\cite{bennaim1999,nie2002}. In three dimensions, 
they differ with $\theta_{BA}^{mf}=6/5$ and $\theta_{BE}=3/2$. 
However, simulations that measure $\theta$ in three dimensions have been
inconclusive, being limited by small system sizes and times, and the measured value of 
$\theta$ ranges from  $\theta= 1.35-1.6$~\cite{Schen2000} to
$\theta \sim 1$~\cite{luding1,luding2}. 
Thus, it remains an open question as to which of the theories, if either, is correct.

In this paper, we study the freely cooling granular gas in three dimensions 
using event-driven molecular dynamics simulations and 
conclude that $\theta \approx \theta_{BA}^{mf}$, conclusively ruling
out $\theta_{BE}$ as a possible solution. Comparing with the results of 
three dimensional BA, we find that $\theta_{BA}^{mf}$ describes the
energy decay in BA only when densities are high and multi particle collisions 
are dominant. We also find that the cluster size and
the velocity distributions of the particles in the granular gas and BA
are strikingly different from each other.

Consider $N$ identical hard-sphere particles distributed uniformly 
within a periodic three-dimensional box of linear length $L$ and with
initial velocities chosen from a normal distribution. The mass 
and diameter of the particles are set equal to $1$. 
All lengths, masses and times are measured in units of particle diameter,
particle mass, and initial mean collision time.
The system evolves in time without any external input of 
energy. All particles move ballistically until they undergo momentum
conserving, deterministic collisions with other particles:
if the velocities before and after 
collision are ${\bf u}_1$, ${\bf u}_2$, and ${\bf v}_1$, ${\bf v}_2$ 
respectively, then
\begin{equation}
{\bf v}_{1,2}={\bf u}_{1,2}-\frac{1+r}{2} [{\bf 
n}.({\bf u}_{1,2} -{\bf u}_{2,1})]  {\bf n},
\label{eq:collision}
\end{equation}
where $0<r<1$ is the coefficient of restitution and ${\bf n}$ is the 
unit vector directed from the center of particle $1$ to the center of 
particle $2$. Equation~(\ref{eq:collision}) leaves the tangential component of 
the relative velocity unchanged, and reduces the magnitude of the longitudinal 
component by a factor $r$.

The above system is studied using large scale event-driven molecular dynamics 
simulations~\cite{rapaportbook} for system sizes up to $N= 8 \times 10^6$. 
For constant coefficient of
restitution, infinite collisions occur in finite time~\cite{Mcnamara}. 
An efficient scheme of avoiding this 
computational difficulty is to make the collisions elastic ($r = 1$) when the 
relative velocity is less than a cutoff velocity 
$\delta$, and $r=r_0<1$ otherwise~\cite{bennaim1999}.

\begin{figure}
\includegraphics[width=\columnwidth]{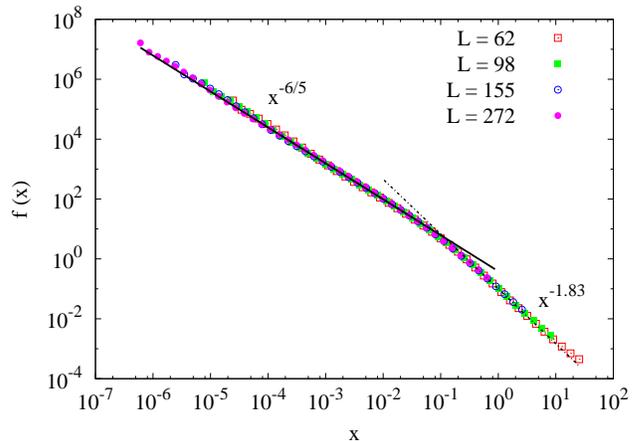}
\caption{\label{fig:engyfseffect} (Color online) The data for kinetic energy
$E(t)$ for different system sizes $L$ collapse onto a single curve
when $t$ and
$E(t)$ are scaled as in 
Eq.~(\ref{Eq:fseffect1}) with $\theta=\theta_{BA}^{mf}=6/5$ and
$z=z_{BA}^{mf}=5/2$. The power law fits are shown by straight lines.
The data are for $\phi=0.208$, $r_0=0.1$, and $\delta=10^{-4}$. 
}
\end{figure}

We first present results for the decrease of kinetic energy with time. 
We find that for $r_{0}=0.10$ and volume fraction $\phi=0.208$, the homogeneous
regime is very short-lived and the inhomogeneous regime is reached at early
times. However, the energy decay deviates from the universal power law $t^{-\theta}$ 
for times larger than a crossover time that increases with 
system size $L$.   We assume that $E(t)$ obeys the finite size scaling form 
\be
E(t) \simeq L^{-z \theta}f\left(\frac{t}{L^{z}}\right),~~t, L \rightarrow
\infty,
\label{Eq:fseffect1}
\ee
where $z$ is the dynamical exponent, and the
scaling function $f(x) \sim x^{-\theta}$ for $x=t L^{-z} \ll 1$.
The simulation data 
for different $L$ collapse onto a single curve (see
Fig.~\ref{fig:engyfseffect}) when 
$E(t)$ and $t$ are
scaled as in Eq.~(\ref{Eq:fseffect1}) with $\theta=\theta_{BA}^{mf}=6/5$ and
$z=z_{BA}^{mf}=5/2$. The power law $x^{-6/5}$ extends over nearly 5
decades, confirming that the energy decay in the 
freely cooling granular gas in three
dimensions has the exponents that are numerically indistinguishable
from the mean-field BA. The data conclusively rules out
$\theta_{BE}=3/2$ as being the correct exponent.
From Fig.~\ref{fig:engyfseffect}, we see that
$f(x) \sim x^{-\eta}$ for $x \gg 1$ with 
$\eta \approx 1.83$, such that at large times $t \gg L^z$,
$E(t) \sim L^{1.58} t^{-1.83}$.

\begin{figure}
\includegraphics[width=\columnwidth]{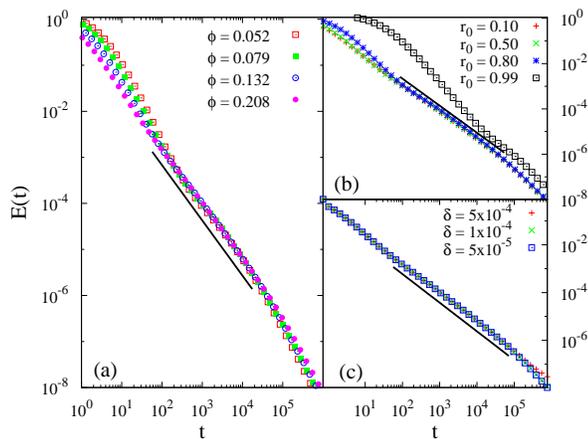}
\caption{\label{fig:engyden} (Color online) The dependence of kinetic
energy $E(t)$ on (a)volume fraction $\phi$, (b) the coefficient of
restitution $r_0$, and (c) the parameter $\delta$. The solid lines are
power laws $t^{-6/5}$. The data is for $\phi=0.208$, $r_{0}=0.10$, $\delta=10^{-4}$
unless it is the varying parameter.}
\end{figure}

We now show that $\theta$ measured from the data in
Fig.~\ref{fig:engyfseffect} is independent of the volume fraction
$\phi$, coefficient of restitution $r_0$, and
$\delta$. The systems with varying $\phi$ are prepared by fixing $L=272$ 
and varying $N$ from $2\times 10^{6} (\phi = 0.052)$ to $8\times 10^{6} 
(\phi = 0.208)$. With increasing $\phi$, we find that the
crossover from homogeneous ($E(t) \sim t^{-2}$) to inhomogeneous
regime ($E(t) \sim t^{-6/5}$) occurs at earlier times [see
Fig.~\ref{fig:engyden}(a)]. In the
inhomogeneous regime, the curves are indistinguishable from each
other. Thus, we see that the exponent $\theta=6/5$ holds even in the limit
$\phi \rightarrow 0$.
Similarly, with increasing $r_0$, though the inhomogeneous regime sets
in at later times, it nevertheless exists with the same power law
$t^{-\theta}$ [see Fig.~\ref{fig:engyden}(b)]. Similar behavior has
been observed in one and two dimensions~\cite{bennaim1999,nie2002}.
We also find no discernible dependence of the data on the
parameter $\delta$ [see Fig.~\ref{fig:engyden}(c)]. Finally, we check
that using a more realistic velocity dependent coefficient of
restitution does not change the value of the exponent
$\theta$ (see supplementary
material~\cite{supplementary}).

We note that $\theta_{BA}^{mf}$ need not be equal to the actual BA
exponent $\theta_{BA}$~\cite{trizac1,trizac4}. We study this
discrepancy in three dimensions by simulating BA directly.
Two colliding particles are replaced with a single
particle whose volume is the sum of the volumes of the colliding
particles. The newly formed aggregate may overlap with other particles
leading to a chain of aggregation events. These multi-particle
collisions result in the exponent $\theta_{BA}$ being dependent on the
volume fraction $\phi$. We find that as $\phi$ increases from $0.005$
to $0.208$, $\theta_{BA}$ decreases from $1.283 \pm 0.005$ to $1.206
\pm 0.005$ and appears to converge to the $\theta_{BA}^{mf}=1.2$ 
with increasing $\phi$. Thus, it is remarkable that the mean field
result describes well only the systems with 
$\phi \gtrsim 0.2$, while
its derivation~\cite{carnevale} assumes the limit $\phi \rightarrow 0$.
\begin{figure}
\includegraphics[width=\columnwidth]{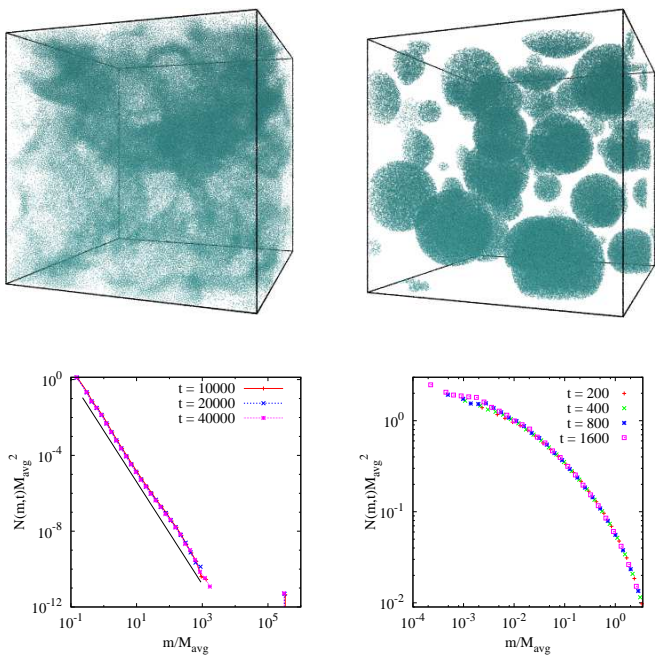}
\caption{\label{fig:snapshots} (Color online)
Snapshots of granular gas (upper left) and BA (upper right)
in the inhomogeneous regime. The lower panel shows the scaled mass
distribution for the granular gas (left) and BA (right).
$M_{\rm{avg}}$ is the mean cluster size.  The solid
line is a power law $m^{-2.70}$.
The data are for $\phi=0.208$, $r_0=0.10$.  }
\end{figure}

The energy decay in  granular gas and BA at higher densities 
being similar, how do  
other statistical properties compare? 
We first study clusters of particles in the inhomogeneous regime.
Snapshots of granular gas and BA (see Fig.~\ref{fig:snapshots}) 
show that clusters in granular gas are extended as 
opposed to compact spherical clusters (by construction) in BA. The spatial
distribution of particles is partially quantified by measuring the cluster size distribution 
$N(m,t)$. For
the granular gas, the simulation box is divided into boxes of side
equal to diameter of a particle. A box is said to be occupied if it
contains the center of a particle. Two occupied boxes belong to the
same cluster if connected by nearest neighbor occupied boxes.
$N(m,t)$ for the granular gas and BA, 
shown in the lower panel of Fig.~\ref{fig:snapshots}, are 
significantly different from one another. For the granular gas, $N(m,t)$ consists of two 
parts: a power law ($\sim m^{-2.7}$) and a peak at large cluster sizes. The power law
describes all clusters other than the largest cluster that accounts for the peak. The
largest cluster contains about $75\%$ of the
particles. For BA, $N(m,t)$ is a power law for small cluster sizes ($\sim m^{-0.2}$) and
exponential for cluster sizes larger than the mean cluster size. Both of these distributions
are different from the mean field result for $N(m,t)$ obtained from the Smoluchowski
equation describing the temporal evolution of $N(m,t)$:
\bea
&&\dot{ N}(m,t)=  \sum_{m_1=1}^{m-1}
\!\! N(m_1,t) N(m\!-\!m_1,t) K(m_1, m-m_1) \nonumber\\
&& \mbox{}- 
2 \sum_{m_1=1}^\infty N(m_1,t) N(m,t) K(m_1,m) ~ m=1,2,\ldots,
\label{eq:smoluchowski}
\eea
where 
\be
K(m_1,m_2) \propto (m_1^{-1/2} + m_2^{-1/2}) (m_1^{1/3}+m_2^{1/3})^2
\label{eq:kernel}
\ee
is
the collision kernel~\cite{leyvraz2005,ccareview}. For this kernel, it
is known that that $N(m,t) \sim \exp(-const \times m^{-1/2})$ for small $m$ 
and $N(m,t) \sim \exp(-const \times m)$ for large
$m$~\cite{leyvraz2005}. While the simulation results for BA matches
for large $m$, it is different (being a power law) for small $m$.
\begin{figure}
\includegraphics[width=\columnwidth]{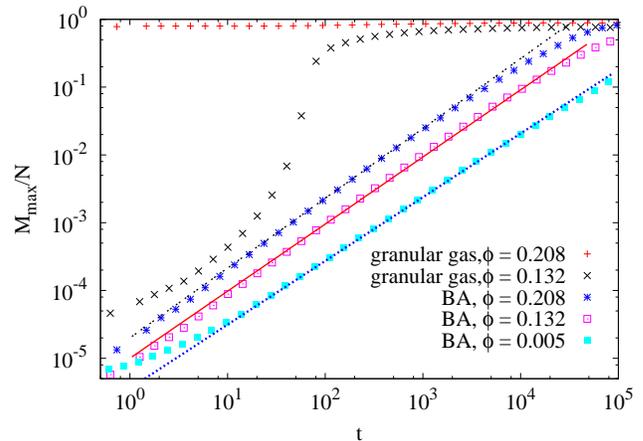}
\caption{\label{fig:largestmass}(Color online) The largest mass $M_{\mathrm{max}}$ as a
function of time. For the granular gas, $r_0 = 0.10$, $\delta=10^{-4}$.
Straight lines are power laws $t^{0.94}$, $t^{0.99}$, $t^{1.03}$ (bottom to top).
}
\end{figure}

Also, for the kernel in Eq.~(\ref{eq:kernel}), 
it is expected that the largest cluster size increases with time $t$ as 
a power law $t^{6/5}$~\cite{leyvraz2005,ccareview}, the mean
field answer. We compare this prediction with the simulations for the
granular gas and BA.
For the granular gas,
rather than a power law growth as in one and two dimensions and in mean field, 
there is
a rapid increase in $M_{\textrm{max}}$ (see upper two curves of
Fig.~\ref{fig:largestmass}) at a time that coincides
with the onset of the inhomogeneous cooling regime. This rapid growth is 
similar to the gelation transition where a gel containing a fraction
of the total number of particles is formed in finite time.  However 
the kernel for BA is non-gelling with mass dimension $1/6$, whereas
the gelation transition requires mass dimension to be larger than $1$~\cite{leyvraz2005,ccareview}.
For BA, $M_{\textrm{max}}$ increases as a power law (see bottom
three curves of
Fig.~\ref{fig:largestmass}), with an exponent that
increases with $\phi$, and possibly converges to the mean field value $6/5$.
Similar behavior is seen for the growth of average cluster size of
BA which grows as a power law with an exponent ranging from 
$1.06$ for $\phi=0.005$ to 
$1.19$ for $\phi=0.313$.

\begin{figure}
\includegraphics[width=\columnwidth] {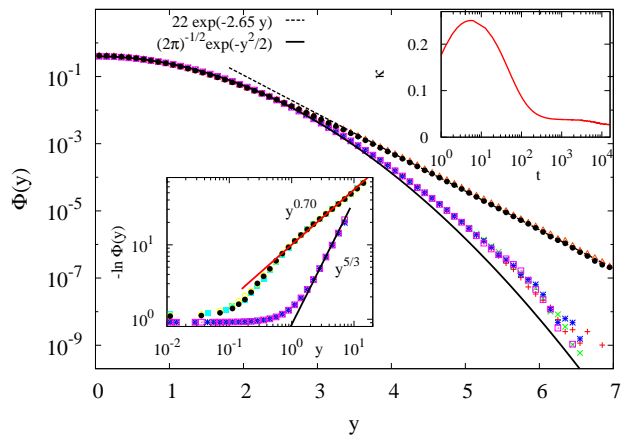}
\caption{\label{fig:veldist}(Color online) The scaled velocity distribution 
function $\Phi(y)$  for the granular gas at times $t = 5,10$ (upper collapsed data) and 
$t= 2000, 4000, 6000, 8000$ (lower collapsed
data). The solid curve is a Gaussian.
The data are for $\phi=0.208$, $r_0 = 0.10$.  Upper inset: The
kurtosis $\kappa$ as a function of time $t$.
Lower inset: $-\ln \Phi(y)$ as a function of $y$ for the granular gas
(lower data) and BA (upper data).  
For BA, the times are $t=400, 800, 1600$ and $\phi=0.208$. 
}
\end{figure}

We further compare the velocity distributions $P(v,t)$, where $v$ is
any velocity component, of the granular gas with that of BA.
$P(v,t)$ has the scaling form $P(v,t) = v_{rms}^{-1} \Phi(v/v_{rms})$,
where $v_{rms}$ is the time dependent root mean square velocity. The
scaling function $\Phi(y)$ is shown in 
Fig.~\ref{fig:veldist} for different times. For the granular gas, at short 
times when the system is homogeneous ($t=5, 10$ in
Fig.~\ref{fig:veldist}), $\Phi(y)$
is an exponential $e^{-\alpha y}$ as predicted by kinetic theory.
We find $\alpha = 2.65$, in good agreement with the kinetic 
theory value 2.60~\cite{ernst}. For larger times  ($t=2000$ --
$8000$ in Fig.~\ref{fig:veldist}), $\Phi(y)$ is
clearly non-Gaussian and has a tail that is overpopulated compared to
the Gaussian (see comparison with Gaussian in Fig.~\ref{fig:veldist}). 
A quantitative measure of the 
deviation from the Gaussian is the kurtosis, 
$\kappa = \langle v^{4} \rangle/\langle v^{2} \rangle ^2 - 5/3$, shown
in the upper inset of Fig.~\ref{fig:veldist}. The kurtosis after an
initial increase,  decreases and saturates to a non-zero value, 
showing quantitatively
that $\Phi(y)$ is non-Gaussian. 
The large $y$ behavior of $\Phi(y)$ is shown 
in the bottom inset of
Fig.~\ref{fig:veldist}. It has been  argued, based on the
probability that a particle never undergoes a collision  up to time $t$, 
that $-\ln[\Phi(y)] \sim y^{2/\theta}$, $y \gg
1$~\cite{nie2002}. For the granular gas, we find that $-\ln[\Phi(y)] \sim y^{5/3}$,
consistent with $\theta=6/5$. However, for  BA we find
$-\ln[\Phi(y)] \sim y^{0.70}$. This is surprising because the
argument that leads to $-\ln[\Phi(y)] \sim
y^{2/\theta}$~\cite{nie2002} is quite general and does not depend on
the detailed dynamics. Thus, in addition to having different
velocity distributions, the argument based on survival
probability fails for BA.

To summarize, we showed that the energy $E(t)$ of a three dimensional
freely cooling granular gas decreases as $t^{-\theta}$, with $\theta\approx
6/5$, indistinguishable from the mean field result for dilute ballistic
aggregation. This rules out Burgers like equations as a description of
the granular gas at large times. We also showed that the relation to
ballistic aggregation appears coincidental with the energy of
the dilute ballistic
gas decaying with a different exponent. In addition, the cluster size
distribution as well as the velocity distribution of ballistic
aggregation  are strikingly
different from that of the granular gas. We hope that this study will
prompt research into finding the correct continuum equations for the
granular gas as well as in the design of experiments to probe the
inhomogeneous cooling regime.  While frictionless 
freely cooling experiments have been limited to the homogeneous
regime, inhomogeneous clustering has
been observed in granular cooling experiments where
only one particle or location
is excited~\cite{boudet2009,jaeger2008,toussaint}. For these systems, scaling 
arguments based on the sticky gas explain the experimental 
results~\cite{zahera,sudhir_dae,sudhir2012}. Multiple localized
excitations may result in a crossover to the freely cooling system,
making such experiments more suitable to probing the inhomogeneous
regime.

\begin{acknowledgments}
The simulations were carried out on the
supercomputing
machine Annapurna at The Institute of Mathematical Sciences.
\end{acknowledgments}

\end{document}